\documentclass[twocolumn]{article}
\RequirePackage{authblk}
\RequirePackage{graphicx}
\RequirePackage{commath}
\RequirePackage{esint}
\begin{document}

\title{System of excited monopole-antimonopole pair in the Weinberg-Salam model}
\author{Dan Zhu, Khai-Ming Wong, Guo-Quan Wong}
\affil{\textit{School of Physics, Universiti Sains Malaysia, 11800 USM Penang}}
\maketitle

\begin{abstract}
We investigate further the properties of axially symmetric monopole-antimonopole pair in the standard Weinberg-Salam model. By using a novel data sampling approach, we have obtained and analyzed 300 numerical solutions corresponding to physical Higgs self-coupling $\beta=0.7782$ and the Weinberg angle $\tan\theta_{\scalebox{.5}{\mbox{W}}}=0.5358$. We calculate the energy of these solutions and confirm that they reside in a range of 13.1690 - 21.0221 TeV. In addition, a unique pattern is shown when the data are arranged according to an algorithm based on the system's symmetry, which seems to indicate the system is oscillating. We also calculate numerically the magnetic charge of the solutions and confirm that their values are indeed $\pm\frac{4\pi}{e}\sin^2\theta_{\scalebox{.5}{\mbox{W}}}$.
\end{abstract}

\section{Introduction}
\label{intro}
Since P.A.M. Dirac \cite{Dirac} introduced magnetic monopoles into Maxwell's theory, it became a subject of extensive studies. Wu and Yang \cite{WuYang} generalized the idea to non-Abelian gauge theories in 1968. However, both Dirac and Wu-Yang monopole possess infinite energy due to the presence of point singularity. The first magnetic monopole solution that possesses finite energy was the 't Hooft-Polyakov monopole \cite{tHooftPolyakov} in SU(2) Yang-Mills-Higgs (YMH) theory as the gauge potential is well-defined in all space. The mass of 't Hooft-Polyakov monopole was estimated to be of order 137 $m_{\scalebox{.5}{\mbox{W}}}$, where $m_{\scalebox{.5}{\mbox{W}}}$ is the mass of intermediate vector boson.

Exact monopole \cite{PrasadSommerfield} and multimonopole \cite{ExactMulti} solutions exist only in the BPS limit with vanishing Higgs potential. Outside the BPS limit only numerical solutions are known.  There also exists non-BPS solutions, such as the axially symmetric monopole-antimonopole pair (MAP) of Kleihaus and Kunz \cite{KKMAP} and monopole-antimonopole chain (MAC) of Kleihaus et al. \cite{KKMAC}. These solutions do not satisfy the Bogomol'nyi equation but only the second order equations, even in the limit of vanishing Higgs potential. They possess finite energy and represent a chain of magnetic monopoles lying in alternating order along the symmetrical axis. These configurations are parameterized by $\theta$-winding number $m(>1)$ and  $\phi$-winding number $n=1$ and $n=2$.

Unfortunately, the SU(2) YMH theory is unrealistic. Solutions found in the SU(2)$\times$U(1) Weinberg-Salam model would possess more important physics implications. The pioneering work was done by Cho and Maison \cite{ChoMaison}, the solution found is known as the electroweak monopole or simply Cho-Maison monopole, a hybrid between Dirac monopole and 't Hooft-Polyakov monopole. The Cho-Maison monopole describes a real monopole through the physical W-boson and Higgs field. Although the Cho-Maison monopole has a singularity at the origin which renders the energy divergent, it has been shown that it is possible to regularize the solution and the mass was estimated to be around 4 to 10 TeV \cite{ChoMaisonMass}. More recently, a different method to regularize the energy has been reported \cite{ChoMaisonRegularization}, suggesting the new BPS bound for Cho-Maison monopole may not be smaller than 2.98 TeV, more probably 3.75 TeV. After the discovery of Higgs boson, electroweak monopole stands as the final test for the Standard Model. For this reason, experimental
detectors around the globe are actively searching for it \cite{Detection1,Detection2,Detection3,Detection4}.

Another line of research in SU(2)$\times$U(1) Weinberg-Salam theory includes the work of Y. Nambu \cite{Nambu} that demonstrates the existence of a pair of magnetic monopole and antimonopole bound by a flux string. The total energy of this MAP configuration is finite and the mass of the system (monopole, antimonopole together with the string) is estimated to be in TeV range. At asymptotically large distances, the real electromagetic field is a linear combination of U(1) and SU(2) gauge fields. Although the arguments and calculations were not rigorous, the existence of such string-like configurations attracts a great deal of interest.

After the work of Nambu \cite{Nambu}, there are more works done on the classical solutions of Weinberg-Salam theory. A well-known case is the sphaleron (sphaleron is static, particle-like, localized in space and unstable solution of the Weinberg-Salam field equations). Klinhamer and Manton in their work \cite{HamerManton} first coined the term `sphaleron' and shown that it possesses baryon number $Q_{\scalebox{.5}{\mbox{B}}}=\frac{1}{2}$. They also notice that there is an electric current in the U(1) field. On the other hand, the works of Hindmarsh and James \cite{MarshJames}, and Radu and Volkov \cite{RaduVolkov} also found that within the sphaleron there is a monopole-antimonopole pair and loop of electromagnetic current. There are also some other works \cite{AxialSymmetry1,AxialSymmetry2} on axially symmetric sphaleron system that uses a similar but more general ansatz, which shows system of sphalerons lying along the symmetrical axis. Though possessing axial symmetry, these results do not actually reveal the internal structure of sphaleron.

Recently, using the axially symmetric generalized ansatz of that in Ref. \cite{ChoMaison}, configurations of MAP, MAC and vortex rings have been constructed in the standard Weinberg-Salam theory. It was shown explicitly that in the 1-MAP configurations, the pair is bound by a $Z^0$ field flux string and there is an electromagnetic current loop encircling them. This confirms the findings of Nambu \cite{Nambu} years ago. Using a less rigourous calculation, it was also shown that the monopole and antimonopole possess magnetic charge $\pm\frac{4\pi}{e}\sin^2\theta_{\scalebox{.5}{\mbox{W}}}$. The 1-MAP in Ref. \cite{Teh} could essentially be viewed as the electroweak counterpart of 1-MAP solution reported by Kleihaus and Kunz \cite{KKMAP} in SU(2) YMH theory. It is also shown in Ref. \cite{Teh} that the 1-MAP is a sphaleron with baryon number $Q_{\scalebox{.5}{\mbox{B}}}=\frac{n}{2}$ and the 2-MAP is a sphaleron anti-sphaleron pair with baryon number $Q_{\scalebox{.5}{\mbox{B}}}=0$, therefore confirming the findings of Ref. \cite{HamerManton,MarshJames,RaduVolkov}.

The solutions in Ref. \cite{Teh}, however, only consider (unphysical) Weinberg angle $\theta_{\scalebox{.5}{\mbox{W}}}=\pi/4$ for ease of calculation and the dimension of quantities investigated was not taken into consideration. For this reason, the work done in Ref. \cite{Teh} lacks physical significance. Therefore in this research, to address the issue, an appropriate dimensionless transformation is first performed on the model and then, the data are obtained specifically for physical Higgs self-coupling $\beta = 0.7782$ and Weinberg angle $\tan \theta_{\scalebox{.5}{\mbox{W}}}=0.5358$. These parameters are calculated according to the latest data published by the Particle Data Group \cite{PDG}, which correspond to $m_{\scalebox{.5}{\mbox{W}}}=80.379$ GeV, $m_{\scalebox{.5}{\mbox{Z}}}=91.1876$ GeV and $m_{\scalebox{.5}{\mbox{H}}}=125.10$ GeV. The energy of the 1-MAP solutions thus found is confirmed to reside in a range of 13.1690 - 21.0221 TeV. Additionally, it is important to note that due to the nature of 1-MAP configurations in Weinberg-Salam model, the conventional method used in obtaining numerical data has limitations. All data were obtained using a novel data sampling approach.

Furthermore, even if the model used in this research is static, the obtained data manifests a unique pattern, which seems to indicate the 1-MAP configuration in the Weinberg-Salam model possesses time-dependent features. Such a pattern is ubiquitously present in all aspects of the solution. Of course, it must be emphasized here that the solutions reported in this paper are obtained from a static model. One must solve the time dependent equations to decisively confirm said claim, i.e., functions that depend on $t, r$ and $\theta$, but the emergence of such solutions even from a time-independent model is interesting to warrant in-depth investigation. Finally, we also calculate numerically the magnetic charge of the solutions and confirm that their values are indeed $\pm\frac{4\pi}{e}\sin^2\theta_{\scalebox{.5}{\mbox{W}}}$, as predicted by Y. Nambu \cite{Nambu}.

The paper is organized as follows. In Section 2 we briefly present the SU(2)$\times$U(1) Weinberg-Salam theory. The numerical method used to acquire and analyse the solutions will be discussed in Section 3. This includes the axially symmetric ansatz, dimensionless transformation, total energy and energy density, unitary gauge, electromagnetic and neutral fields, and the novel data sampling approach. The 1-MAP confiurations at physical value of Higgs self-coupling $\beta=0.7782$ and Weinberg angle $\tan\theta_{\scalebox{.5}{\mbox{W}}}=0.5358$ are analyzed and discussed in Section 4. The scope of this study includes the energy, magnetic properties and the unique pattern of the 1-MAP configuration in Weinberg-Salam model. We end with some comments in Section 5.

\section{Standard Weinberg-Salam Model}
\label{sec:2}
The Lagrangian of the standard Weinberg-Salam model is given by \cite{ChoMaison,Teh}
	\begin{align}
	\mathcal{L}=&-\left(\mathcal{D}_\mu\pmb\phi\right)^\dagger\left(\mathcal{D}^\mu\pmb\phi\right)-\frac{\lambda}{2}\left(\pmb\phi^\dagger\pmb\phi-\frac{\mu_{\scalebox{.5}{\mbox{H}}}^2}{\lambda}\right)^2\nonumber\\
	&-\frac{1}{4}F^a_{\mu\nu}F^{a\mu\nu}-\frac{1}{4}G_{\mu\nu}G^{\mu\nu}.
	\label{eqn:Lagrangian}
	\end{align}
Here, $\mathcal{D}_\mu$ is the covariant derivative of SU(2)$\times$U(1) group, which is defined as
	\begin{equation}
	\mathcal{D}_\mu=D_\mu-\frac{i}{2}g'B_\mu=\partial_\mu-\frac{i}{2}gA^a_\mu\sigma_a-\frac{i}{2}g'B_\mu,
	\end{equation}
where $D_\mu$ is the covariant derivative of SU(2) group only.\par
The SU(2) gauge coupling constant, potential and electromagnetic tensor are given by $g$, $A^a_\mu$ and $F^a_{\mu\nu}$. Their counterparts in U(1) are denoted by $g'$, $B_\mu$ and $G_{\mu\nu}$. The term, $\sigma_a$, is the Pauli matrices, $\pmb\phi$ and $\lambda$ are the complex scalar Higgs doublet and Higgs field self-coupling constant. Higgs boson mass and $\mu_{\scalebox{.5}{\mbox{H}}}$ are related through $m_{\scalebox{.5}{\mbox{H}}}=\sqrt2\mu_{\scalebox{.5}{\mbox{H}}}$. In addition, the Higgs field can be expressed as $\pmb\phi=H\pmb\xi/\sqrt{2}$, where $\pmb\xi$ is a column 2-vector that satisfies $\pmb\xi^\dagger\pmb\xi=1$, to further simplify the calculation. Metric used in this paper is ($-$+++).\par
Through Lagrangian (\ref{eqn:Lagrangian}), three equations of motion can be obtained as the following,
	\begin{align}
	\mathcal{D}^\mu\mathcal{D}_\mu\pmb\phi&=\lambda\left(\pmb\phi^\dagger\pmb\phi-\frac{\mu_{\scalebox{.5}{\mbox{H}}}^2}{\lambda}\right)\pmb\phi,\nonumber\\
	D^\mu F^a_{\mu\nu}&=\frac{ig}{2}\left[\pmb\phi^\dagger\sigma^a\left(\mathcal{D}_\nu\pmb\phi\right)-\left(\mathcal{D}_\nu\pmb\phi\right)^\dagger\sigma^a\pmb\phi\right],\nonumber\\
	\partial^\mu G_{\mu\nu}&=\frac{ig'}{2}\left[\pmb\phi^\dagger\left(\mathcal{D}_\nu\pmb\phi\right)-\left(\mathcal{D}_\nu\pmb\phi\right)^\dagger\pmb\phi\right].\label{eqn:EoM}
	\end{align}

\section{The Numerical Method}
\label{sec:3}

\subsection{Axially Symmetric Mangetic Ansatz}
\label{sec:3-1}
To obtain the electrically neutral 1-MAP solutions in the Weinberg-Salam model, the following magnetic ansatz is employed \cite{Teh}:
	\begin{align}
	gA^a_i&=-\frac{1}{r}\psi_1(r,\theta)\hat{n}^a_\phi\hat{\theta}_i+\frac{n}{r}\psi_2(r,\theta)\hat{n}^a_\theta\hat{\phi}_i\nonumber\\
	&\quad\,+\frac{1}{r}R_1(r,\theta)\hat{n}^a_\phi\hat{r}_i-\frac{n}{r}R_2(r,\theta)\hat{n}^a_r\hat{\phi}_i,\nonumber\\
	g'B_i&=\frac{n}{r}B_1(r,\theta)\hat{\phi}_i,\;gA^a_0=g'B_0=0,\nonumber\\
	\Phi^a&=\Phi_1(r,\theta)\hat{n}^a_r+\Phi_2(r,\theta)\hat{n}^a_\theta=\frac{H(r,\theta)}{\sqrt2}\hat{\Phi}^a,\label{eqn:Ansatz}
	\end{align}
where the unit vector, $\hat{\Phi}^a$, can be expressed as,
	\begin{align}
	\hat{\Phi}^a&=-\pmb\xi^\dagger\sigma^a\pmb\xi=\cos(\alpha-\theta)\hat{n}^a_r+\sin(\alpha-\theta)\hat{n}^a_\theta,\nonumber\\
	\pmb\xi&=i
		\begin{pmatrix}
		\sin\frac{\alpha(r,\theta)}{2}e^{-in\phi}\\
		-\cos\frac{\alpha(r,\theta)}{2}
		\end{pmatrix}.
	\end{align}
In magnetic ansatz (\ref{eqn:Ansatz}), the spatial spherical coordinate unit vectors are
	\begin{align}
	\hat{r}_i&=\sin\theta\cos\phi\,\delta_{i1}+\sin\theta\sin\phi\,\delta_{i2}+\cos\theta\,\delta_{i3},\nonumber\\
	\hat{\theta}_i&=\cos\theta\cos\phi\,\delta_{i1}+\cos\theta\sin\phi\,\delta_{i2}-\sin\theta\,\delta_{i3},\nonumber\\
	\hat{\phi}_i&=-\sin\phi\,\delta_{i1}+\cos\phi\,\delta_{i2}, 
	\end{align}
whereas the unit vectors for isospin coordinate system are given by 
	\begin{align}
	\hat{n}^a_r&=\sin\theta\cos n\phi\,\delta^a_1+\sin\theta\sin n\phi\,\delta^a_2+\cos\theta\,\delta^a_3,\nonumber\\
	\hat{n}^a_\theta&=\cos\theta\cos n\phi\,\delta^a_1+\cos\theta\sin n\phi\,\delta^a_2-\sin\theta\,\delta^a_3,\nonumber\\
	\hat{n}^a_\phi&=-\sin n\phi\,\delta^a_1+\cos n\phi\,\delta^a_2. 
	\end{align}
The $\phi$-winding number, $n$, is set to 1. The angle, $\alpha(r,\theta)$ approaches $p\theta$ asymptotically \cite{Teh}, where $p$ is the parameter controlling the number of poles in the solution, which is set to two for 1-MAP solutions.

\subsection{Dimensionless Transformation}
\label{sec:3-2}
Upon substituting magnetic ansatz (\ref{eqn:Ansatz}) into the equations of motion (\ref{eqn:EoM}), the system of equations was reduced to seven coupled second-order partial differential equations, which were further simplified with the following substitutions,
	\begin{equation}
	x=m_{\scalebox{.5}{\mbox{W}}}r,\;\widetilde{H}=\frac{H}{H_0},\;\tan\theta_{\scalebox{.5}{\mbox{W}}}=\frac{g'}{g},\;\beta^2=\frac{\lambda}{g^2}.
	\end{equation}
Here, $H_0=\sqrt2\mu_{\scalebox{.5}{\mbox{H}}}/\sqrt\lambda$ and $m_{\scalebox{.5}{\mbox{W}}}=gH_0/2$. The new radial coordinate, $x$, is dimensionless, which is then compactified through $\widetilde{x}=x/(x+1)$.

Using finite difference approximation method, the set of seven coupled equations of motion was converted into a system of non-linear equations, which was then discretized onto a non-equidistant grid of $M\times N$, where $M=70$, $N=60$. The region of integration covers all space which translates to $0\leq\widetilde{x}\leq1$ and $0\leq\theta\leq\pi$. The rescaled Higgs field, $\widetilde{H}$, approaches one asymptotically.

Only two parameters were left after the transformation, $\tan\theta_{\scalebox{.5}{\mbox{W}}}$ and $\beta$, which were calculated to be 0.5358 and 0.7782 according to the latest data published by the Particle Data Group \cite{PDG}.

\subsection{The Energy}
\label{sec:3-3}
In the standard SU(2)$\times$U(1) Weinberg-Salam model, the energy density for 1-MAP configuration is obtained from the energy-momentum tensor, $T^{\mu\nu}$:
	\begin{align}
	\varepsilon=T_{00}&=\frac{1}{4}G_{ij}G_{ij}+\frac{1}{2}G_{i0}G_{i0}+\frac{1}{4}F^a_{ij}F^a_{ij}+\frac{1}{2}F^a_{i0}F^a_{i0}\nonumber\\
	&\quad\,+\left(\mathcal{D}_i\pmb\phi\right)^\dagger\left(\mathcal{D}_i\pmb\phi\right)+\left(\mathcal{D}_0\pmb\phi\right)^\dagger\left(\mathcal{D}_0\pmb\phi\right)\nonumber\\
	&\quad\,+\frac{\lambda}{2}\left(\pmb\phi^\dagger\pmb\phi-\frac{\mu_{\scalebox{.5}{\mbox{H}}}^2}{\lambda}\right)^2.
	\end{align}
However, as magnetic ansatz (\ref{eqn:Ansatz}) is electrically neutral, all temporal components vanish and only four terms are left in the above expression:
	\begin{align}
	\varepsilon&=\frac{1}{4}G_{ij}G_{ij}+\frac{1}{4}F^a_{ij}F^a_{ij}+\left(\mathcal{D}_i\pmb\phi\right)^\dagger\left(\mathcal{D}_i\pmb\phi\right)\nonumber\\
	&\quad\,+\frac{\lambda}{2}\left(\pmb\phi^\dagger\pmb\phi-\frac{\mu_{\scalebox{.5}{\mbox{H}}}^2}{\lambda}\right)^2,\label{eqn:epsilon}
	\end{align}
with
	\begin{align}
	\varepsilon_{\scalebox{.5}{\mbox{U(1)}}}&=\frac{1}{4}G_{ij}G_{ij},\;\varepsilon_{\scalebox{.5}{\mbox{SU(2)}}}=\frac{1}{4}F^a_{ij}F^a_{ij},\nonumber\\
	\varepsilon_{\scalebox{.5}{\mbox{H}}}&=\left(\mathcal{D}_i\pmb\phi\right)^\dagger\left(\mathcal{D}_i\pmb\phi\right)+\frac{\lambda}{2}\left(\pmb\phi^\dagger\pmb\phi-\frac{\mu_{\scalebox{.5}{\mbox{H}}}^2}{\lambda}\right)^2,
	\end{align}
representing the contributions from U(1), SU(2) gauge fields and Higgs field respectively.

Now, after performing the dimensionless transformation, the $\varepsilon$ in equation (\ref{eqn:epsilon}) becomes $(m_{\scalebox{.5}{\mbox{W}}}^2H_0^2)\widetilde\varepsilon$ where $m_{\scalebox{.5}{\mbox{W}}}^2H_0^2$ carries its dimension and $\widetilde\varepsilon$ is dimensionless. Therefore, the total energy,
	\begin{align}
	E&=\iint r^2\sin\theta\,\varepsilon\,dr\,d\theta\int d\phi\nonumber\\
	&=2\pi\iint\left(\frac{x^2}{m_{\scalebox{.5}{\mbox{W}}}^2}\right)\sin\theta\left(m_{\scalebox{.5}{\mbox{W}}}^2H_0^2\widetilde\varepsilon\right)\,\frac{dx}{m_{\scalebox{.5}{\mbox{W}}}}\,d\theta\nonumber\\
	&=\frac{2\pi H_0^2}{m_{\scalebox{.5}{\mbox{W}}}}\iint x^2\sin\theta\,\widetilde\varepsilon\,dx\,d\theta.\label{eqn:DfulE}
	\end{align}
The double integral is dimensionless and the factor outside of the integral is found to be
	\begin{align}
	\frac{2\pi H_0^2}{m_{\scalebox{.5}{\mbox{W}}}}&=\frac{2\pi\frac{m_{\scalebox{.5}{\mbox{H}}}^2}{\beta^2 g^2}}{m_{\scalebox{.5}{\mbox{W}}}}=\frac{2\pi m_{\scalebox{.5}{\mbox{H}}}^2\sin^2\theta_{\scalebox{.5}{\mbox{W}}}}{m_{\scalebox{.5}{\mbox{W}}}\beta^2 e^2}\nonumber\\
	&=\frac{2\times3.1416\times125.10^2\times0.2230}{80.379\times0.7782^2\times0.303^2}\nonumber\\
	&\approx4.9067\text{ TeV}\label{eqn:factor}
	\end{align}
in natural unit.

\subsection{The Unitary Gauge}
\label{sec:3-4}
To investigate the magnetic property of the solution, magnetic ansatz (\ref{eqn:Ansatz}) is transformed through the gauge,
	\begin{align}
	G&=-i
		\begin{pmatrix}
		\cos\frac{\alpha}{2}&\sin\frac{\alpha}{2}e^{-in\phi}\\
		\sin\frac{\alpha}{2}e^{in\phi}&-\cos\frac{\alpha}{2}
		\end{pmatrix}\nonumber\\
	&=\cos\frac{-\pi}{2}+i\hat{u}^a_r\sigma^a\sin\frac{-\pi}{2},\nonumber\\
	\hat{u}^a_r&=\sin\frac{\alpha}{2}\cos{n\phi}\,\delta^a_1+\sin\frac{\alpha}{2}\sin{n\phi}\,\delta^a_2+\cos\frac{\alpha}{2}\,\delta^a_3.\label{eqn:gauge}
	\end{align}
The transformed Higgs field and the third component of the gauge potential have the following form,
	\begin{align}
	gA'^3_\mu&=\frac{n}{r}\left(\psi_2h_2-R_2h_1-\frac{1-\cos\alpha}{\sin\theta}\right)\hat\phi_\mu=\frac{A_1}{r}\hat\phi_\mu,\nonumber\\
	\Phi'^a&=\delta^a_3,
	\end{align}
The particular gauge in equation (\ref{eqn:gauge}) was chosen because $gA'^3_\mu$ produced is precisely the gauge potential of 't Hooft electromagnetic tensor \cite{Teh}, $\hat F_{\mu\nu}=\hat\Phi^aF^a_{\mu\nu}-\frac{1}{g}\varepsilon^{abc}\hat\Phi^aD_\mu\hat\Phi^bD_\nu\hat\Phi^c$ \cite{tHooftPolyakov}. For this reason, the U(1) and SU(2) magnetic field can be calculated as
	\begin{align}
	g'B^{\scalebox{.5}{\mbox{U(1)}}}_i&=-\frac{g'}{2}\varepsilon^{ijk}G_{jk}\nonumber\\
	&=-\varepsilon^{ijk}\partial_j\left\{nB_1\sin\theta\right\}\partial_k\phi,\nonumber\\
	gB^{\scalebox{.5}{\mbox{SU(2)}}}_i&=-\frac{g}{2}\varepsilon^{ijk}\hat F_{jk}=\varepsilon^{ijk}\partial_j\left(gA'^3_k\right)\nonumber\\
	&=-\varepsilon^{ijk}\partial_j\left\{A_1\sin\theta\right\}\partial_k\phi,\label{eqn:magnetic_field_definition_1}
	\end{align}
respectively. The magnetic field lines can be constructed by drawing the contour lines of the terms in curly brackets. According to Coleman \cite{Coleman}, there is no unique definition of magnetic field outside of Higgs vacuum. As the definition shown in equations (\ref{eqn:magnetic_field_definition_1}) views the configuration as a point charge from afar and hence, it is not preferable when investigating the 3D magnetic charge density at small $r$. Instead, the following definition of the SU(2) magnetic field \cite{MarshJames} is used 
	\begin{equation}
	gB^{\scalebox{.5}{\mbox{SU(2)}}}_i=gB_i^a\hat\Phi^a=-\frac{g}{2}\varepsilon_{ijk}F^{ajk}\hat\Phi^a.\label{eqn:SU2_magnetic_field_definition_2}
	\end{equation}

\subsection{Electromagnetic and Neutral Fields}
\label{sec:3-5}
The physical electromagnetic field, $\mathcal{A}_\mu$, and neutral field, $\mathcal{Z}_\mu$, are related to the gauge fields, $B_\mu$ and $A'^3_\mu$, through the following matrix,
	\begin{align}
		\begin{pmatrix}
		\mathcal{A}_\mu\\
		\mathcal{Z}_\mu
		\end{pmatrix}&=
		\begin{pmatrix}
		\cos\theta_{\scalebox{.5}{\mbox{W}}}&\sin\theta_{\scalebox{.5}{\mbox{W}}}\\
		-\sin\theta_{\scalebox{.5}{\mbox{W}}}&\cos\theta_{\scalebox{.5}{\mbox{W}}}
		\end{pmatrix}
		\begin{pmatrix}
		B_\mu\\
		A'^3_\mu
		\end{pmatrix}\nonumber\\
	&=\frac{1}{\sqrt{g'^2+g^2}}
		\begin{pmatrix}
		g&g'\\
		-g'&g
		\end{pmatrix}
		\begin{pmatrix}
		B_\mu\\
		A'^3_\mu
		\end{pmatrix}\label{eqn:AZrelation},
	\end{align}
or equivalently,
	\begin{align}
	e\mathcal{A}_\mu&=\left[\cos^2\theta_{\scalebox{.5}{\mbox{W}}}\left(g'B_\mu\right)+\sin^2\theta_{\scalebox{.5}{\mbox{W}}}\left(gA'^3_\mu\right)\right],\nonumber\\
	e\mathcal{Z}_\mu&=\cos\theta_{\scalebox{.5}{\mbox{W}}}\sin\theta_{\scalebox{.5}{\mbox{W}}}\left[-\left(g'B_\mu\right)+\left(gA'^3_\mu\right)\right],\label{eqn:AZmu}
	\end{align}
where $e$ is the unit electric charge, which is related to the gauge coupling constants through
	\begin{align}
	e=\frac{gg'}{\sqrt{g^2+g'^2}}.
	\end{align}

The physical magnetic field, $B_i^{\scalebox{.5}{\mbox{EM}}}$, is then calculated according to the mixing shown in equations (\ref{eqn:AZmu}),
	\begin{align}
	eB_i^{\scalebox{.5}{\mbox{EM}}}=\cos^2\theta_{\scalebox{.5}{\mbox{W}}}\left(g'B^{\scalebox{.5}{\mbox{U(1)}}}_i\right)+\sin^2\theta_{\scalebox{.5}{\mbox{W}}}\left(gB^{\scalebox{.5}{\mbox{SU(2)}}}_i\right),
	\end{align}
and through Gauss's law, the magnetic charge enclosed in a Gaussian surface, $S$, with surface element, $dS^i$, can be obtained using the following integral:
	\begin{equation}
	Q_{\scalebox{.5}{\mbox{M}}(S)}=\oiint_{\scalebox{.5}{\mbox{S}}}B_i^{\scalebox{.5}{\mbox{EM}}}dS^i\label{eqn:Q_M(S)}=\iiint_{\scalebox{.5}{\mbox{V}}}\partial^iB_i^{\scalebox{.5}{\mbox{EM}}}dV.
	\end{equation}
Evidently, $\partial^iB_i^{\scalebox{.5}{\mbox{EM}}}$ is the magnetic charge density, $M$.

For the magnetic charge enclosed in the upper hemisphere, the following Gaussian surface was defined:
	\begin{align}
	S_+=H^2_+\cup D^2_{\scalebox{.5}{\mbox{XY}}},
	\end{align}
where $H^2_+$ is a half sphere above the $xy$-plane and $D^2_{\scalebox{.5}{\mbox{XY}}}$ denotes a disk in the $xy$-plane centered at the origin, both $H^2_+$ and $D^2_{\scalebox{.5}{\mbox{XY}}}$ have the same radius $r\rightarrow\infty$. Taking into account the correct orientation of the surface elements, the magnetic charge enclosed in this Gaussian surface can be calculated as:
	\begin{align}
	Q_{\scalebox{.5}{\mbox{M}}(S_+)}&=\iint_{H^2_+}B_i^{\scalebox{.5}{\mbox{EM}}}r^2\sin\theta\,d\theta\,d\phi\;\hat r^i\nonumber\\
	&\quad\,-\iint_{D^2_{\scalebox{.5}{\mbox{XY}}}}B_i^{\scalebox{.5}{\mbox{EM}}}r\,dr\,d\phi\;\hat z^i.
	\end{align}

\subsection{Data Sampling Approach}
\label{sec:3-6}
In the numerical procedures, good initial guesses are needed for the numerical computations to converge. In the work of Teh et al. \cite{Amin}, an initial guess is chosen to obtain a solution, said solution would be used as the new initial guess to calculate subsequent solution, which is then used as an initial guess again. For instance, the solution with $\beta = 0.1$ is used as the initial guess for $\beta = 0.2$, the solution obtained would then be used as the initial guess for calculating $\beta = 0.3$ and the procedure continues until a desired value of $\beta$ is reached. However, this method is no longer applicable to the 1-MAP configuration in the Weinberg-Salam model (more details in section \ref{sec:4-2}). Therefore, a different approach is adopted in this paper.

The initial guess is a column matrix that is divided into seven equal segments, each corresponds to the seven profile functions set in the magnetic ansatz (\ref{eqn:Ansatz}). A numerical solution of the 1-MAP configuration in the SU(2) YMH theory with $\beta=0.7782$ is converted into six out of the seven segments needed for the initial guess. The remaining segment represents $B_1\left(r,\theta\right)$, which is unique to the Weinberg-Salam model, and is filled with zeros. This completes the construction of a single initial guess. 

When a new initial guess is needed, the previously used 1-MAP solution in the SU(2) YMH theory would be recalculated so that it is slightly different from the one used before. Essentially, they are the same solution in the SU(2) YMH theory, with each entry differs by an amount smaller than $10^{-11}$. However, such minute variations in initial guess resulted in fairly different numerical solutions in the Weinberg-Salam model. In this research, a total of 300 initial guesses were constructed this way and a wide spectrum of numerical solutions were obtained in the Weinberg-Salam model. However, it must be emphasized here that the wide spectrum of solutions stems from the fundamental nature of the 1-MAP configuration in the Weinberg-Salam model. This is because the aforementioned apporach, when applied to the SU(2) YMH model, produces exactly the same result as the method used in Ref. \cite{Amin}.

Moreover, the pole separation, $d_z$, is widely considered as a good indicator of the system's total energy. However, due to the presence of a flux string between the monopole and antimonopole, $d_z$ of the 1-MAP solutions obtained in this research can no longer be accurately measured. Therefore, a novel algorithm is deviced in place of $d_z$. It is designed to quantify the degree of symmetry of Higgs modulus, $\abs{\Phi}=\sqrt{\Phi_1^2+\Phi_2^2}$. We define the difference in symmetry, $D$, for any numerical solution as: 
	\begin{align}
	D&=\frac{\sum_{p=1}^{M}\sum_{q=1}^{N/2}}{\sum_{p=1}^{M}\sum_{q=1}^{N/2}1}\abs{\frac{\abs{\Phi}(p,q)}{\abs{\Phi}(p,N+1-q)}-1}\nonumber\\
	&\quad\,(1\leq p\leq M,\;1\leq q\leq N/2),\label{eqn:algorithm}
	\end{align}
where $(p,\;q)$ is the location of any point in the upper hemisphere and $(p,\;N+1-q)$ is the location of its image in the lower hemisphere. The absolute value then calculates the relative difference of $\abs{\Phi}$ at these two points, which is rounded down to the nearest tenth and maxed out at 1. The summations in the numerator and denominator simply mean the relative difference is summed and averaged over all points in the upper hemisphere. Finally, $D$ is expressed as a percentage value and used as the quantified criterion for all subsequent analyses.

Ideally, when $\abs{\Phi}$ of a particular solution is completely symmetrical about the $xy$-plane, $D=0\%$. In general, the more asymmetrical $\abs{\Phi}$ (about the $xy$-plane) becomes, the larger the value of $D$. Say, if a solution possesses $D=5\%$, it could be interpreted as ``For this particular solution, $\abs{\Phi}$ in the upper hemisphere is 5\% different from $\abs{\Phi}$ in the lower hemisphere, according to algorithm (\ref{eqn:algorithm})."

\section{Results and Discussion}
\label{sec:4}

\subsection{Boundary Conditions}
\label{sec:4-1}
By fixing boundary conditions at small distances ($r\rightarrow0$), large distances ($r\rightarrow\infty$), along positive and negative $z$-axis ($\theta=0$ and $\pi$), the system of seven coupled second-order partial differential equations were solved. Asymptotically, 
	\begin{align}
	\psi_A(\infty,\theta)&=2,R_A(\infty,\theta)=B_1(\infty,\theta)=0,\nonumber\\
	\Phi_1(\infty,\theta)&=\cos\theta,\Phi_2(\infty,\theta)=\sin\theta,
	\label{eqn:BCs}
	\end{align}
where $A=1$, 2. Then, for small distances, 
	\begin{align}
	&\psi_A(0,\theta)=R_A(0,\theta)=B_1(0,\theta)=0,\nonumber\\
	&\Phi_1(0,\theta)\sin\theta+\Phi_2(0,\theta)\cos\theta=0,\nonumber\\
	&\partial_r(\Phi_1(r,\theta)\cos\theta-\Phi_2(r,\theta)\sin\theta)|_{r=0}=0.
	\end{align}
And finally, along positive and negative $z$-axis, 
	\begin{align}
	\partial_\theta\psi_1=\partial_\theta\psi_2=R_1=R_2=\partial_\theta\Phi_1=\Phi_2=B_1=0.\label{eqn:BCe}
	\end{align}
The equations were solved numerically with boundary conditions (\ref{eqn:BCs}) - (\ref{eqn:BCe}).

\subsection{General Properties}
\label{sec:4-2}
\begin{figure}[t]
	\includegraphics[width=\linewidth]{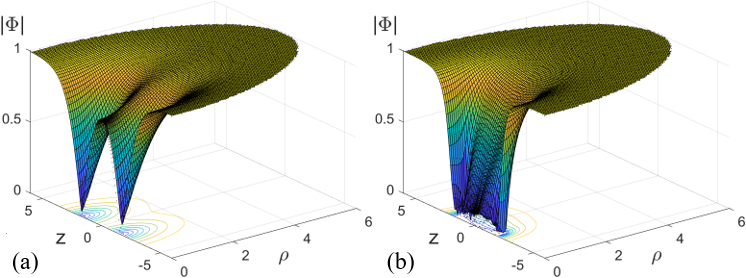}
	\caption{3D Higgs modulus plots for 1-MAP solutions found in (a) SU(2) YMH theory and (b) SU(2)$\times$U(1) Weinberg-Salam model.}
	\label{fig:HiggsCompare}
	\end{figure}

Figure \ref{fig:HiggsCompare} shows a comparison between 3D Higgs modulus of 1-MAP solutions found in different models. 
Configurations in SU(2) YMH theory, such as those reported by Kleihaus and Kunz \cite{KKMAP}, are two separate entities as in Fig. \ref{fig:HiggsCompare}(a) with a well-defined pole separation. In SU(2) $\times$ U(1) Weinberg-Salam model, however, the pair is bound by a flux string as in Fig. \ref{fig:HiggsCompare}(b), the pole separation, $d_z$, can no longer be accurately measured. The $\rho$-axis is defined as $\rho=\sqrt{x^2+y^2}$. It has been proven that the flux string is unstable for a range of Weinberg angle, $0\leq\sin^2\theta_{\scalebox{.5}{\mbox{W}}}<0.8$, when $m_{\scalebox{.5}{\mbox{H}}}>24$ GeV \cite{UnstableString}. Since both conditions are satisfied in reality, the numerical results presented here are all unstable saddle point solutions in the Weinberg-Salam model.

\begin{figure}[t]
	\includegraphics[width=\linewidth]{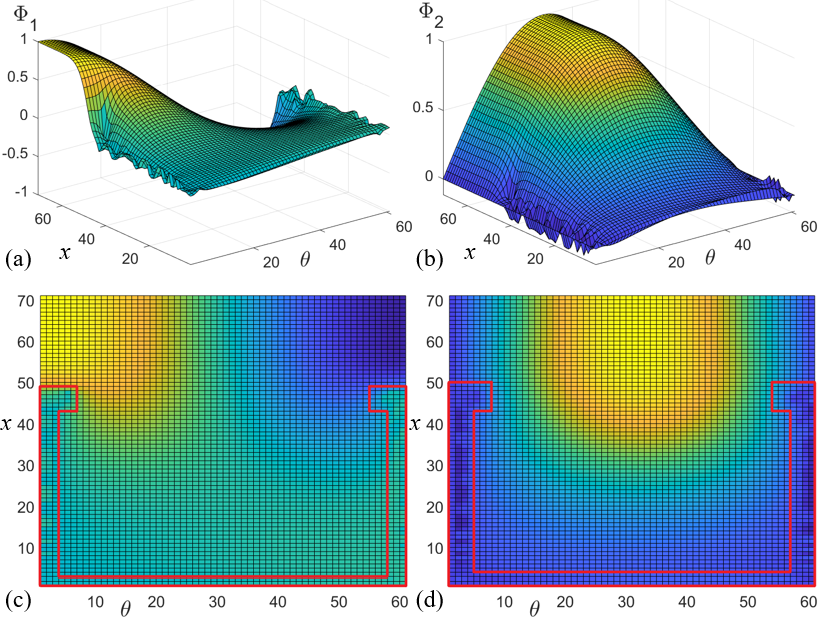}
	\caption{Surface plots of (a) $\Phi_1$ and (b) $\Phi_2$, with corresponding plots viewed from the top for (c) $\Phi_1$ and (d) $\Phi_2$, for the particular solution displayed in Fig. \ref{fig:HiggsCompare}(b).}
	\label{fig:ProfilePhi12}
	\end{figure}

Figure \ref{fig:ProfilePhi12} shows the surface plots of $\Phi_1$ and $\Phi_2$ for the particular solution displayed in Fig. \ref{fig:HiggsCompare}(b). 
Along certain boundaries, the regions that are not smooth in Fig. \ref{fig:ProfilePhi12}(a) and (b) correspond to small $r$ and along $z$-axis, which is precisely where the flux string is located. These areas are outlined in red and can be seen clearly in the top views of respective surface plots as shown in Fig. \ref{fig:ProfilePhi12}(c) and (d). Such irregularities only exist in $\Phi_1$ and $\Phi_2$. However, employing the method used in Ref. \cite{Amin} to obtain solutions would spread the irregularities onto other profile functions, which renders the conventional method unfeasible. Therefore, the aforementioned data sampling approach was devised. 

\subsection{The Energy}
\label{sec:4-3}
\begin{figure*}[t]
	\includegraphics[width=\linewidth]{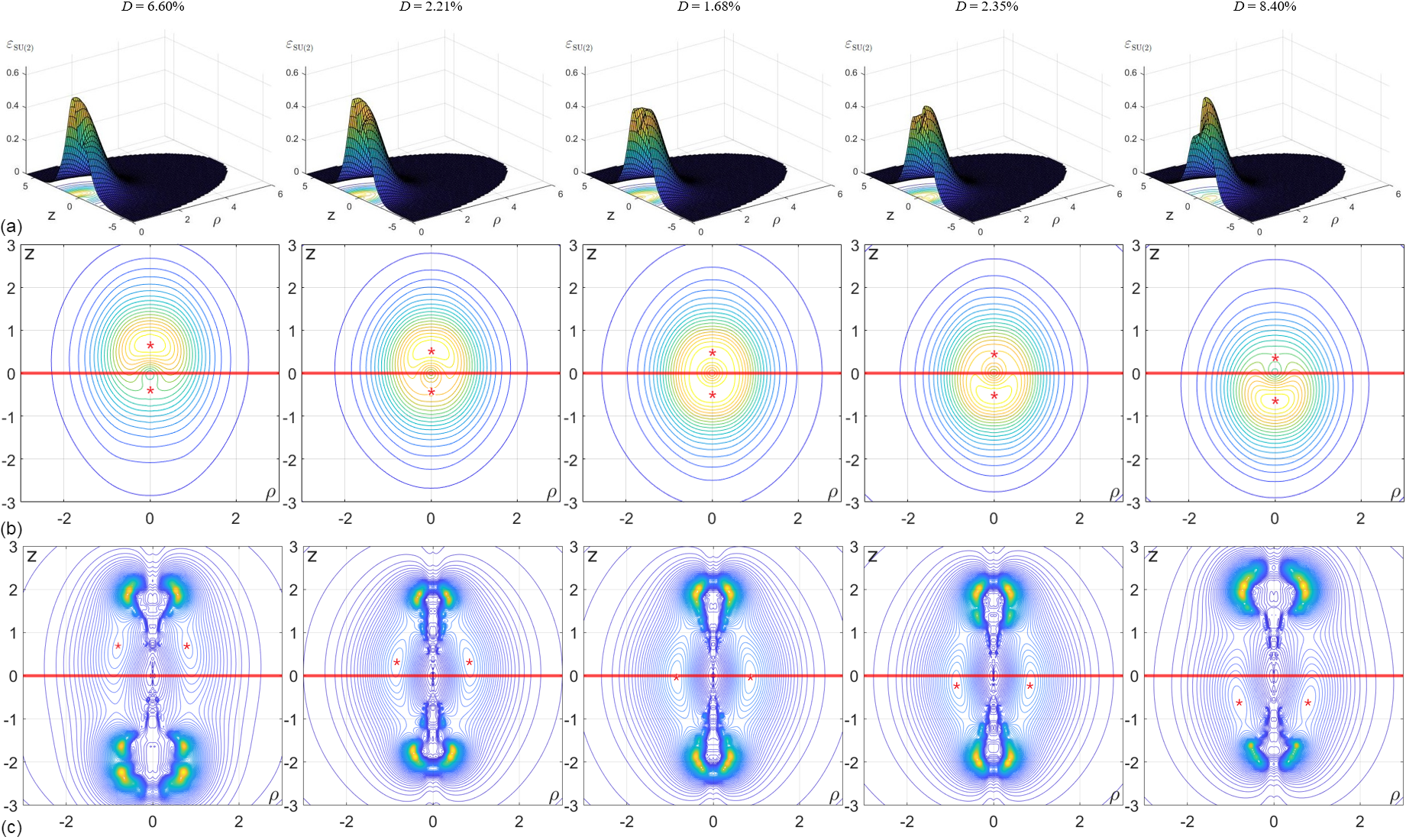}
	\caption{(a) 3D and (b) contour plots of $\varepsilon_{\scalebox{.5}{\mbox{SU(2)}}}$ along with (c) the combined weighted energy density contours of selected solutions.}
	\label{fig:SU2Combined}
	\end{figure*}

The 3D and contour plots for $\varepsilon_{\scalebox{.5}{\mbox{SU(2)}}}$ of selected solutions are shown in Fig. \ref{fig:SU2Combined}(a) and (b).
Using algorithm (\ref{eqn:algorithm}), the value of $D$ for each solution was calculated and displayed alongside of Fig. \ref{fig:SU2Combined}. Highly asymmetrical solutions possessing higher $D$ = 6.60\% and 8.40\% were put on either side of Fig. \ref{fig:SU2Combined}, while the most symmetrical configuration having the lowest $D=1.68\%$ is positioned in the middle. The corresponding 3D and contour plots for $\varepsilon_{\scalebox{.5}{\mbox{U(1)}}}$ possess a similar shape as Fig. \ref{fig:SU2Combined}(a) and (b), albeit with much lower intensity. From the $\varepsilon_{\scalebox{.5}{\mbox{SU(2)}}}$ 3D plots, two peaks can be identified for each solution. Both are located on the $z$-axis with one in the upper hemisphere (positive $z$) and the other in the lower hemisphere (negative $z$). The exact location for the peaks are marked with an asterisk in the corresponding contour plots as shown in Fig. \ref{fig:SU2Combined}(b).

\begin{figure}[t]
	\includegraphics[width=\linewidth]{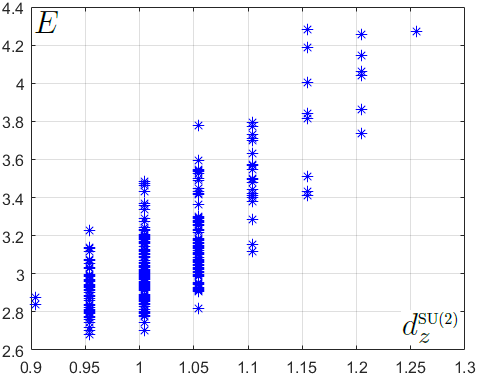}
	\caption{Dimensionless total energy, $E$, versus $d_z^{\scalebox{.5}{\mbox{SU(2)}}}$ for all 300 data collected.}
	\label{fig:E_versus_dz}
	\end{figure}

Now, even if the pole separation, $d_z$, of the 1-MAP solutions found in this research cannot be measured accurately, one can still use the peak separation in $\varepsilon_{\scalebox{.5}{\mbox{SU(2)}}}$ (the distance between asterisks in Fig. \ref{fig:SU2Combined}(b)) as an indicator to where the monopoles are. This separation is referred to as $d_z^{\scalebox{.5}{\mbox{SU(2)}}}$ and the dimensionless total energy, $E$, versus $d_z^{\scalebox{.5}{\mbox{SU(2)}}}$ for all 300 data collected are shown in Fig. \ref{fig:E_versus_dz}.
The horizontal axis is discrete because the readings for the peak location are taken from an interpolated mesh grid. The value ranges from 0.9038 to 1.2552 with a clear pattern that $E$ increases with $d_z^{\scalebox{.5}{\mbox{SU(2)}}}$. This can be explained by the fact that the 1-MAP is bound by a flux string, increasing the pole separation would bring the system to an excited state, therefore increasing the system's total energy. The minimum separation in $\varepsilon_{\scalebox{.5}{\mbox{SU(2)}}}$ is found to be $d_z^{\scalebox{.5}{\mbox{SU(2)}}}=0.9038$. Additionally, it is found that solutions with larger $d_z^{\scalebox{.5}{\mbox{SU(2)}}}$ also have higher $D$. Thus, asymmetrical solutions have higher peak separation in $\varepsilon_{\scalebox{.5}{\mbox{SU(2)}}}$.

Figure \ref{fig:SU2Combined}(c) shows the contours of combined weighted energy density, $(r^2\sin\theta)\varepsilon$. Due to the $1/r^2$ factors present in the expression for $\varepsilon_{\scalebox{.5}{\mbox{H}}}$, profile functions need to be accurate in order for $\varepsilon_{\scalebox{.5}{\mbox{H}}}$ not to blow up at small $r$. The irregularities in $\Phi_1$ and $\Phi_2$ therefore created fake singularities near the origin and along the $z$-axis. For this reason, the combined weighted energy density was plotted instead, which includes contributions from all three sources. In the figure, there exists a bump (marked with asterisks). It corresponds to a region with higher energy density and mainly comes from $\varepsilon_{\scalebox{.5}{\mbox{SU(2)}}}$. The $\sin\theta$ factor in weighted energy density expression is necessary as the bump is otherwise unobservable. Furthermore, the position of the bump is closely related to the system's symmetry and always confined to the string connecting the monopole and antimonopole. When the system is symmetric (lower $D$), the bump is always located at $z=0$. 

\begin{figure*}[t]
	\includegraphics[width=\linewidth]{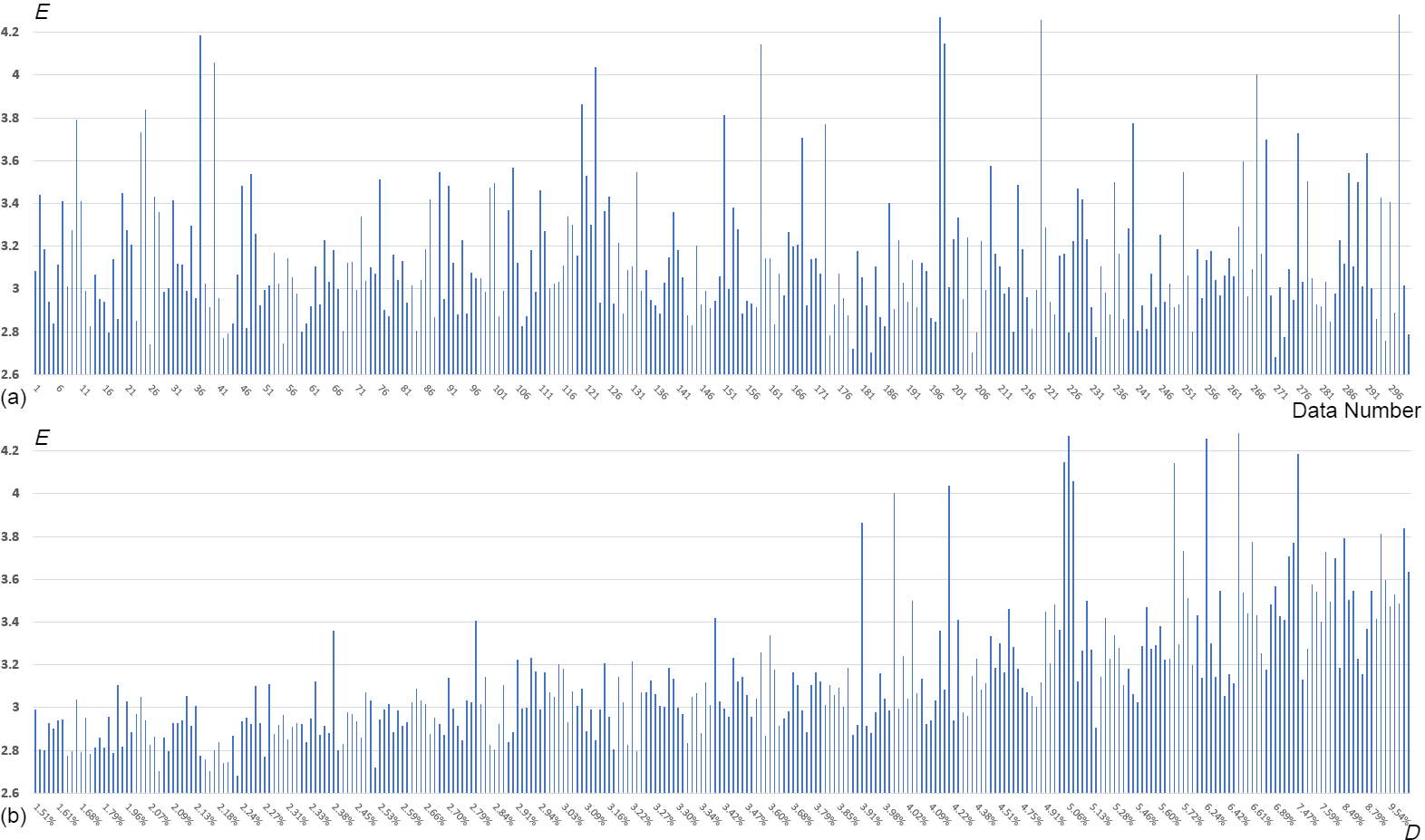}
	\caption{Dimensionless $E$ of the system versus (a) data number, (b) difference in symmetry, $D$.}
	\label{fig:0p7782}
	\end{figure*}

The total energy of the 1-MAP system is finite because mathematically, gauge potentials defined in magnetic ansatz (\ref{eqn:Ansatz}) are well-defined in all space. Figure \ref{fig:0p7782}(a) shows the dimensionless $E$ for all 300 data when $\beta=0.7782$, $\tan{\theta_{\scalebox{.5}{\mbox{W}}}}=0.5358$ in the Weinberg-Salam model, arranged according to the order by which they are obtained. 
The dimensionless $E$ ranges from 2.6834 to 4.2844 with no apparent pattern at first sight. However, it becomes clear when the data were arranged according to $D$ from low to high as shown in Fig. \ref{fig:0p7782}(b). Higher energy spikes all shifted to the right where solutions become more and more asymmetrical. System with collectively higher $D$ tends to possess higher energy than system with symmetry. It must be stressed here that a system could have relatively low energy even with higher $D$, but this occurrence is less frequent.

Finally, multiplying the dimensionless $E$ with the factor shown in equation (\ref{eqn:factor}) and taking all 300 data into consideration, we found the 1-MAP configuration in the SU(2)$\times$U(1) Weinberg-Salam model resides in a range of 13.1690 - 21.0221 TeV.

\subsection{Magnetic Properties}
\label{sec:4-4}
\begin{figure*}[t]
	\includegraphics[width=\linewidth]{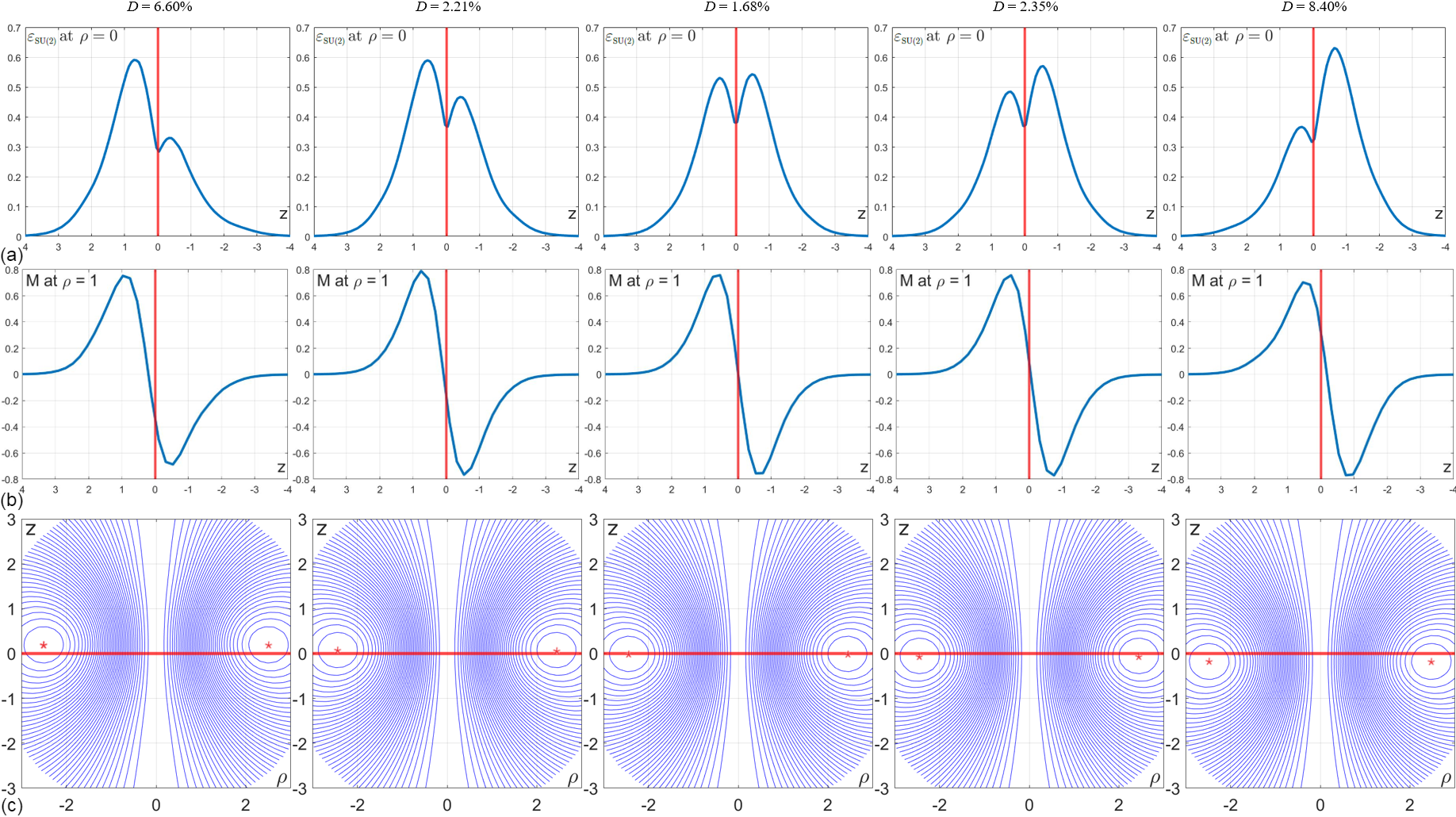}
	\caption{(a) Cross sections of $\varepsilon_{\scalebox{.5}{\mbox{SU(2)}}}$ at $\rho=0$, (b) $M$ at $\rho=1$ and (c) U(1) magnetic field lines of the data shown in Fig. \ref{fig:SU2Combined}.}
	\label{fig:CS}
	\end{figure*}

Similar to $\varepsilon$, the irregularities in $\Phi_1$ and $\Phi_2$ created fake singularities near the origin when plotting the 3D magnetic charge density plots. In this case, the cross sections of magnetic charge density, $M$, at $\rho=1$ were plotted and shown in Fig. \ref{fig:CS}(b). The definition of $M$ takes the mixing in equation (\ref{eqn:AZmu}) into account, with the U(1) part defined in equation (\ref{eqn:magnetic_field_definition_1}) and SU(2) part shown in equation (\ref{eqn:SU2_magnetic_field_definition_2}). However, the U(1) part is calculated to be zero and the contribution is solely from the SU(2) gauge field.

Monopole with positive magnetic charge is located in the upper hemisphere (positive $z$), while antimonopole with negative charge is in the lower hemisphere as seen in Fig. \ref{fig:CS}(b). The system's symmetry is also well-reflected by our algorithm. The solution in the middle of Fig. \ref{fig:CS}(b) has the lowest $D$ and possesses near complete symmetry in magnetic charge distribution, whereas for configurations with higher $D$, the system would shift to either hemisphere. One thing to note is that both poles shift in the same direction. More specifically, the positive peak located at ($z$, $M$) = (0.9697, 0.7531) for the solution with $D=6.60\%$ in Fig. \ref{fig:CS}(b) gradually shifts to (0.5387, 0.7030) for the one with $D=8.40\%$.

The total magnetic charge of the system is zero, however, the charge carried by either pole was predicted to be $\frac{4\pi}{e}\sin^2\theta_{\scalebox{.5}{\mbox{W}}}$ by Y. Nambu \cite{Nambu}. To verify this, we first calculated the magnetic charge in the U(1) gauge field, $Q^{\scalebox{.5}{\mbox{U(1)}}}_{\scalebox{.5}{\mbox{M}}}$. The result is $Q^{\scalebox{.5}{\mbox{U(1)}}}_{\scalebox{.5}{\mbox{M}}}=0.4591\times10^{-7}$, which is averaged over all 300 data collected and indicates that the total magnetic charge in the U(1) gauge field is close to zero. We then move on to $Q^{\scalebox{.5}{\mbox{U(1)}}}_{\scalebox{.5}{\mbox{M}}}$ in the upper hemisphere and find it to be $Q^{\scalebox{.5}{\mbox{U(1)}}}_{\scalebox{.5}{\mbox{M}}(S_+)}=-3.3053\times10^{-6}$, which implies the U(1) magnetic charge in either hemisphere is also close to zero.

This indicates that the 1-MAP does not reside in U(1) and the magnetic charge is solely contributed by the SU(2) gauge field. This is further confirmed by U(1) magnetic field lines plots in Fig. \ref{fig:CS}(c). In addition, judging by the concentric field lines, there must exist a sourceless electric current going into and out of the page at $\rho=\pm2.5$ (marked with asterisks), which is induced by the shift shown in Fig. \ref{fig:CS}(b) according to Faraday's law. The electric current moves in accordance with the shift in magnetic charge distribution.

\begin{figure*}[t]
	\includegraphics[width=\linewidth]{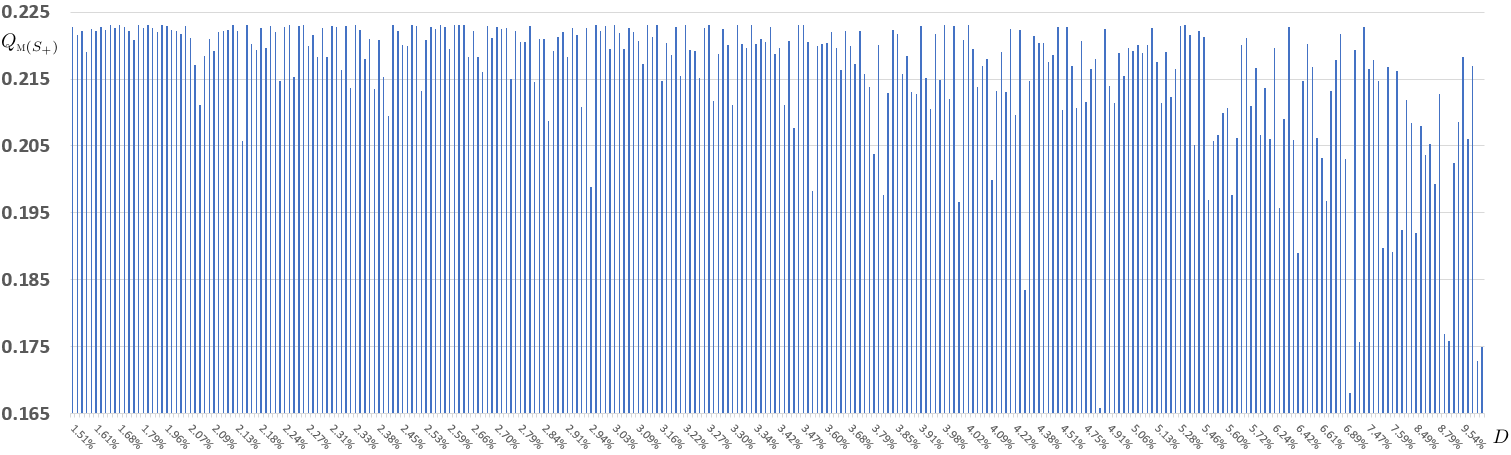}
	\caption{Magnetic charge in the upper hemisphere, $Q_{\scalebox{.5}{\mbox{M}}(S_+)}$, arranged according to $D$ from low to high for all 300 data.}
	\label{fig:TotalM}
	\end{figure*}

In unit of $4\pi/e$, Nambu's prediction has the value of $\sin^2\theta_{\scalebox{.5}{\mbox{W}}}\approx0.2230$. Figure \ref{fig:TotalM} shows the magnetic charge enclosed in the upper hemisphere, $Q_{\scalebox{.5}{\mbox{M}}(S_+)}$, calculated from all solutions obtained, arranged according to $D$ from low to high. 
In Fig. \ref{fig:TotalM}, the vast majority of solutions obtained matches the predicted value. Significant drops in $Q_{\scalebox{.5}{\mbox{M}}(S_+)}$ appear more frequently in asymmetrical solutions (Higher $D$). As shown in Fig. \ref{fig:CS}(b), $M$ does not distribute evenly across hemispheres in asymmetrical solutions. This indicates that some negative magnetic charge is present in the Gaussian surface, $S_+$, which leads to a lower amount of the total magnetic charge enclosed, therefore resulting in a smaller value of $Q_{\scalebox{.5}{\mbox{M}}(S_+)}$.

\subsection{The Oscillating Pattern}
\label{sec:4-5}
In Fig. \ref{fig:SU2Combined} and \ref{fig:CS}, there exists a unique pattern that seems to indicate the solutions are oscillating, even if the model used in this research is static. In Fig. \ref{fig:SU2Combined}(a), the highest value in $\varepsilon_{\scalebox{.5}{\mbox{SU(2)}}}$ 3D plots gradually shifts downwards along the $z$-axis. In addition, the cross sections of $\varepsilon_{\scalebox{.5}{\mbox{SU(2)}}}$ at $\rho=0$ is shown in Fig. \ref{fig:CS}(a) to provide a clearer view of this process. Likewise, the weighted energy density bump (indicated by asterisks) in Fig. \ref{fig:SU2Combined}(c) moves from the upper hemisphere, across the red line, to the lower hemisphere, forming a clear trajectory. Similarly, Fig. \ref{fig:CS}(b) depicts the continuous process of positive magnetic charge moving towards the lower hemisphere and in Fig. \ref{fig:CS}(c), the location of the electric current (indicated by asterisks) in the U(1) gauge field also changes accordingly. The movement in magnetic charge distribution and U(1) electric current contribute directly to the movement in energy density.

\begin{figure*}[t]
	\includegraphics[width=\linewidth]{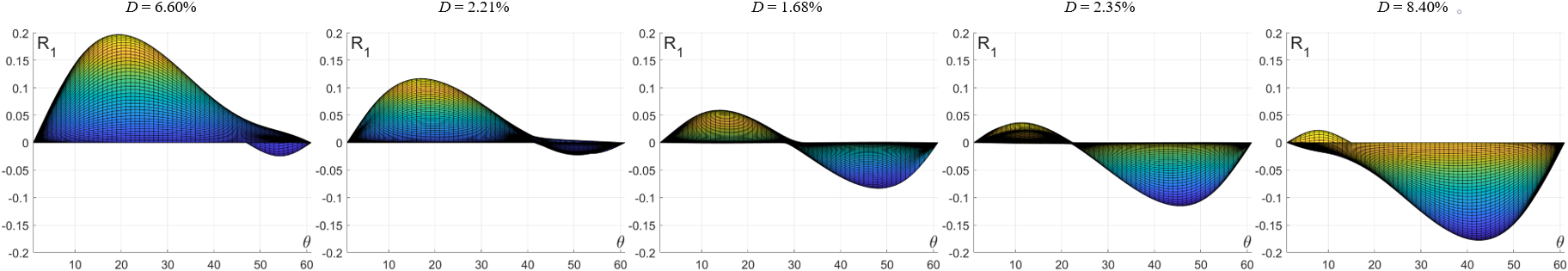}
	\caption{$R_1$-$\theta$ view of surface plots of profile function $R_1$ for the solutions shown in Fig. \ref{fig:SU2Combined}.}
	\label{fig:R1}
	\end{figure*}

Such an oscillating pattern is manifested in multiple aspects of the solution. This is because, fundamentally, profile functions behave this way. The $R_1$-$\theta$ view of the surface plots of profile function $R_1$ is shown in Fig. \ref{fig:R1} in which $R_1$ oscillates.
This kind of oscillation is ubiquitously present in all seven profile functions, which directly leads to oscillation in all aspects of the configuration.

\begin{figure*}[t]
	\includegraphics[width=\linewidth]{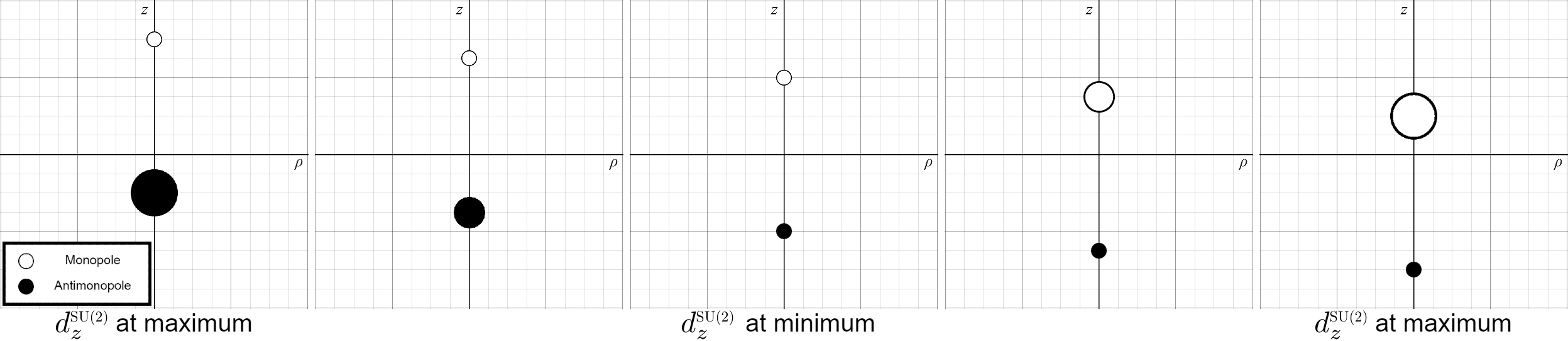}
	\caption{An illustration of the 1-MAP movement pattern}
	\label{fig:illustration}
	\end{figure*}

Additionally, it is worth noting that in Fig. \ref{fig:SU2Combined}(c), when the weighted energy density bump is in the upper hemisphere, the antimonopole in the lower hemisphere visibly becomes energetically excited and vice versa. Now, since both $\varepsilon_{\scalebox{.5}{\mbox{U(1)}}}$ and $\varepsilon_{\scalebox{.5}{\mbox{SU(2)}}}$ behave similarly, this means when the weighted energy density bump is in the upper hemisphere, $\varepsilon_{\scalebox{.5}{\mbox{H}}}$ is higher in the lower hemisphere or in other words, the oscillation of $\varepsilon_{\scalebox{.5}{\mbox{H}}}$ and oscillations of $\varepsilon_{\scalebox{.5}{\mbox{U(1)}}}$ and $\varepsilon_{\scalebox{.5}{\mbox{SU(2)}}}$ are out of phase. The exact reason causing this is currently unknown, which deserves further investigation. Moreover, as seen in Fig. \ref{fig:CS}(a) and (b), both monopole and antimonopole move in the same direction, which is in direct contrast to a traditional dipole oscillation, where two charges move in opposite directions alternating between repelling and attracting. An illustration of this process is shown in Fig. \ref{fig:illustration}.
Larger solid and hollowed circles depict a more energetically excited state of the monopole and antimonopole. As mentioned in section (\ref{sec:4-3}), $d_z^{\scalebox{.5}{\mbox{SU(2)}}}$ is larger in asymmetrical solutions, this is also shown in Fig. \ref{fig:illustration}.

\begin{figure}[t]
	\includegraphics[width=\linewidth]{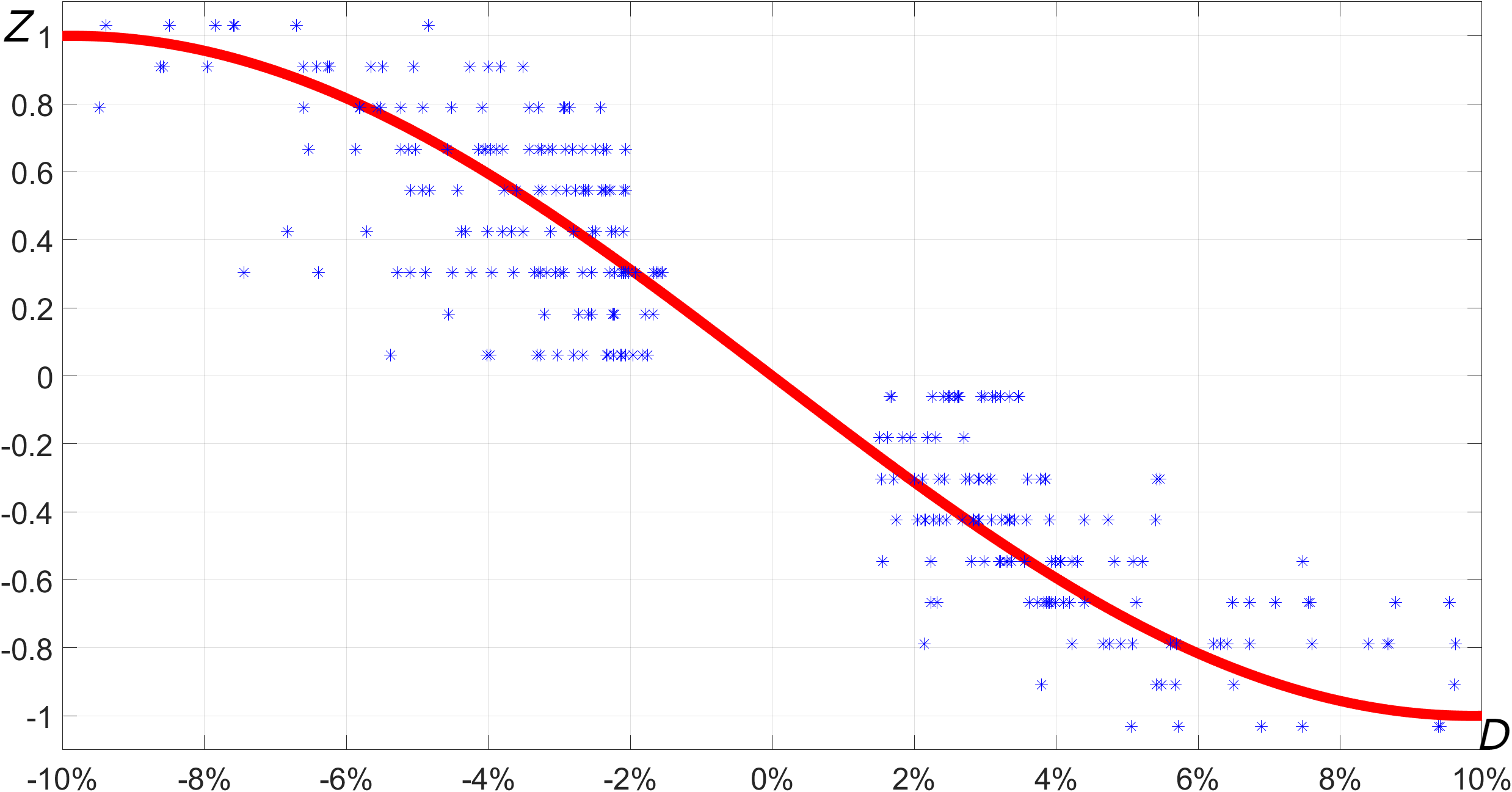}
	\caption{Half-sinusoidal wave pattern fitted from all 300 data collected.}
	\label{fig:Sinusoidal}
	\end{figure}

Lastly, using the location of the weighted energy density bump as an indicator, the following half-sinusoidal curve was fitted from all 300 data collected as shown in Fig. \ref{fig:Sinusoidal}.
The vertical axis  refers to the $z$ coordinate of the peak of the bump in Fig. \ref{fig:SU2Combined}(c). Additionally, when $z$ is positive, $-D$ is used instead in the horizontal axis in Fig. \ref{fig:Sinusoidal}. Clearly, in the vicinity of the fitted curve, data points appear more compacted. The gap in the middle of the plot was the result of no solution with $D$ less than 1.5\% was found. However, it must be stressed here that the model used in this research is static. One must solve the time-dependent equations to describe a complete oscillating behavior and to obtain a full sinusoidal wave pattern.

\section{Conclusion}
\label{sec:conclu}
We have studied numerical solutions of SU(2)$\times$U(1) Weinberg-Salam theory corresponding to axially symmetrical 1-MAP configuration, which is a monopole-antimonopole system bound by a $Z^0$ field flux string with an electric current loop encircling them. By using a novel data sampling approach, 300 distinct solutions are constructed for physical Higgs self-coupling $\beta = 0.7782$ and Weinberg angle $\tan\theta_{\scalebox{.5}{\mbox{W}}}=0.5358$. We calculate the energy of these solutions and find that they reside in a range of 13.1690 - 21.0221 TeV. We also verify numerically the magnetic charge of the solutions and confirm that their values are indeed $\pm\frac{4\pi}{e}\sin^2\theta_{\scalebox{.5}{\mbox{W}}}$, as predicted by Y. Nambu \cite{Nambu}.

The collection of all data, when arranged according to $D$, exhibits a pattern that indicates the system is oscillating. Such a pattern is ubiquitously found in all aspects of the solution because all seven profile functions oscillate. In the weighted energy density contour plots, the bump visibly forms a trajectory. The shift in magnetic charge distribution and the oscillation in U(1) electric current contribute directly to the movement in energy density. It is also worth noting that the oscillation of $\varepsilon_{\scalebox{.5}{\mbox{H}}}$ and oscillations of $\varepsilon_{\scalebox{.5}{\mbox{U(1)}}}$ and $\varepsilon_{\scalebox{.5}{\mbox{SU(2)}}}$ are out of phase, when the bump is in the upper hemisphere, the antimonopole in the lower hemisphere becomes energetically excited and vice versa. The origin of this phenomenon is unknown and deserves more study in the future.

Furthermore, the magnetic charges do not move in the conventional manner, where the dipole would become infinitesimally close and then distanced up to a maximum separation, back and forth. However here, both magnetic charges shift in the same direction along the $z$-axis. In addition, besides encircling the 1-MAP system, the position of the electric current in the U(1) gauge field also changes in accordance with the general oscillating pattern.

On a final note, the observation of oscillating solutions derived from a time-independent system is unexpected. Of course, there remains numerous difficulties and limitations when inferring the behaviors of a dynamic solution using a static model. For example, we cannot tell the direction in which the bump is moving, nor the rate at which the process is taking place, as well as only half of a sinusoidal wave pattern could be constructed. We argue that the behaviour of the 1-MAP configuration reported here could only be fully understood by solving the time dependent system, i.e., equations that depend on $t,r$ and $\theta$. This pursue will be reported in a later work.

The work done here could also be extended by adding electric charges into the model, forming dyons. It would be interesting to know how the electric charge from the dyon would interact with the U(1) electric current reported here, considering both are electric in nature. Doing so would inevitably produce configurations with even higher masses, but a more complete picture of the 1-MAP configuration could be revealed. We will report these findings in a separate paper.

\end{document}